# Relationship between the level of mental fatigue induced by a prolonged cognitive task and the degree of balance disturbance

Frédéric Noé[1*], Betty Hachard[1], Hadrien Ceyte[2], Noëlle Bru[3], Thierry Paillard[1]

[1] Université de Pau et des Pays de l'Adour / E2S UPPA, MEPS, Tarbes, France

[2] Université de Lorraine, DevAH, 54000, Nancy, France

[3] Université de Pau et des Pays de l'Adour / E2S UPPA, CNRS, LMAP, Anglet, France

[*]**Corresponding author**:

Frédéric Noé, Université de Pau et des Pays de l'Adour / E2S UPPA, MEPS, 11 rue Morane Saulnier, 65000 Tarbes, France

Tel: +33(0)5 62566128

E-mail: frederic.noe@univ-pau.fr

**Abstract**

This study investigated the effects of mental fatigue (MF) induced by a 90-minute AX-continuous performance test (AX-CPT) on balance control by addressing the issue of the heterogeneity of individuals' responses. Twenty healthy young active participants were recruited. They had to carry out two balance tasks (sway as little as possible on a stable support with the eyes open and closed) when standing on a force platform before and after performing a 90-minute AX-CPT. The NASA-TLX test was used to assess the subjective manifestations of MF. Objective cognitive performance was measured using results from the AX-CPT. Inter-individual differences in behavioural deterioration due to MF were analysed with a hierarchical cluster analysis, which categorizes participants' behaviors into subgroups with similar characteristics. The cluster analysis revealed that the achievement of the AX-CPT induced various levels of MF and balance impairments within the whole sample. A significant relationship between the level of MF and the degree of balance disturbance was observed only when participants stood with the eyes open, thus suggesting that inter-individual differences in vulnerability to MF could stem from differences between subjects in the level of engagement of visual attention and/or from differences in field dependency for balance control. These findings show that the completion of the same prolonged demanding cognitive task induces a strong heterogeneity in subjects' responses, with marked individual differences in MF vulnerability that affect balance control differently according to the sensory context.

# 1- Introduction

Mental fatigue (MF) refers to a change in psychophysiological state caused by prolonged period of demanding activity (Boksem et al., 2005; Martin et al., 2018) or sleep deprivation (Angus and Heslegrave, 1985; Fortier-Brochu et al., 2010; Ma et al., 2009). Its symptoms are broad and varied such as an increased feeling of tiredness, an increased resistance against further effort, a lack of energy, and/or a decreased motivation and alertness (Martin et al., 2018; Van Cutsem et al., 2017). Numerous studies showed that MF could impair cognitive performances and affect executive functions, as illustrated by increased reaction time and/or decreased response accuracy in various cognitive tasks such as stroop task, psychomotor vigilance task or AX-Continuous Performance Test (AX-CPT) (Boksem et al., 2005; Marcora et al., 2009; Pageaux et al., 2013; Smith et al., 2019). MF also negatively influences motor performances, especially in endurance activities where it increases perception of effort thus leading to an early exercise disengagement (Martin et al., 2018; Marcora et al., 2009; Pageaux et al., 2013). MF might also impair balance, a fundamental and essential motor skill in daily living, likely due to impairments of executive functions, reduced attentional resources and disturbances in sensory integration (Hachard et al., 2020; Ma et al., 2009; Morris and Christie, 2020; Varas-Diaz et al., 2020).

Further studies about the effects of MF induced by sleep deprivation have shown that there were large inter-individual differences in neurobehavioral responses to MF (Van Dogen et al., 2004; Tkachenko and Dinges, 2018). Some individuals show no sign of impairment in response to total sleep deprivation or chronic partial sleep restriction whereas others exhibit substantial decreases in vigilance and attentional performance as well as higher-order executive functions (Tkachenko and Dinges, 2018). Similarly, one can assume that the fulfilment of a prolonged cognitive task induces different levels of MF between participants and thus influences balance control differently within a sample of participants with similar characteristics. No study, to our knowledge, has addressed the issue of heterogeneity between participants' response to MF induced by a demanding cognitive task and its effects on balance control.

The purpose of the current study was therefore to investigate differences in individual responses of participants who had to perform the same 90-minute AX-CPT cognitive task. We re-examined our previously published dataset initially analyzed with a standard group ensemble statistical average method (Hachard et al., 2020) in order to 1) form a classification system of MF levels and balance control impairments using a hierarchical cluster analysis (HCA); and 2) determine whether balance control impairments were related to participants' MF levels. Cluster analysis is a multivariate statistical method which places individuals into a cluster that contains other individuals with similar characteristics thus enabling the identification of natural groupings that may exist in a population (White and McNair 2002). It was hypothesized that participants who performed the same prolonged continuous demanding cognitive task would not demonstrate a similar level of MF thus engendering different levels of impairment on balance control.

**2- Materials and methods**

**2-1- Participants**

Twenty healthy young right-handed participants – thirteen men and seven women [mean (SD): age 21.8 (1.7) years old, height: 173.6 (9.8) cm, weight: 69.4 (11.5) kg] volunteered to participate in the study. Exclusion criteria included a documented balance control disorder or a medical condition that might affect balance control, a neurological or mental disorder, proven attention deficit disorder and a musculoskeletal impairment in the past 2 years. Participants were asked to avoid strenuous activity and to maintain regular sleep habits 48h before the data collection session. They were also asked and not to consume caffeine, alcohol, cigarettes or any psychoactive substances on the day of the experiment. All participants voluntarily signed an informed consent form before starting the experiment, which was in accordance with the Helsinki Declaration. All procedures were approved by and performed in accordance with the relevant guidelines and regulations of the University of Pau and Adour Countries Ethics Committee.

**2-2- MF intervention**

A 90-minute AX-CPT, during which participants were required to press the space bar when the letter "X" followed the letter "A" among a series of randomly generated letter sequences, was used to induce MF. The detailed procedure is described in Hachard et al. (2020). This test is a visual cognitive task that requires sustained attention, working memory, response inhibition and error monitoring (Marcora et al., 2009; Pageaux et al., 2013). As illustrated in figure 1, the percentage of correct responses (CR) and reaction time (RT) were calculated during the first and last 10-min periods of the AX-CPT in order to assess the AX-CPT performance at the beginning (PRE) and the end of the MF intervention (POST). The relative differences (RD) of CR and RT (CR_RD and RT_RD) was calculated as follows: RD $=100*[POST-PRE]/PRE$. RD is an easily interpretable descriptor, which characterizes the evolution of the variables between PRE and POST measures while limiting the influence of the heterogeneity between participants at baseline (PRE).

**2-3- Subjective workload assessment**

The National Aeronautics and Space Administration - Task Load Index (NASA-TLX) (Hart and Staveland, 1988), translated and adapted by Maincent et al. (2005), was used to assess subjective mental workload through a paper-and-pencil version. This test was completed after carrying out the 90-minute AX-CPT (Figure 1). It allows for an understanding of the workload sources. This test relies on the following dimensions: mental demand (MD), physical demand (PD), temporal demand (TD), frustration level (FL), effort (EFFORT), performance (PERF) and on a global score (GS). Participants assessed each dimension using non-graduated bipolar scales, ranging from "low" to "high", for which scores were then calculated from 0 to 100. Next, in each paired combination of the six dimensions (totalling fifteen pairs), the subjects chose the dimension which was the most related to the definition of workload, giving a weighting factor for all dimensions. Then, each dimension score was multiplied by the appropriate weight and a global score was calculated by averaging all the weighted dimensions' scores.

**2-4- Balance control assessment**

Balance control was assessed before (PRE) and after (POST) performing the AX-CPT (Figure 1). Participants were asked to stand on a stable force platform (Stabilotest® Techno Concept, Mane, France) and to sway as little as possible for 60 seconds with the eyes open (EO) and closed (EC). In each balance task, participants stood with parallel feet with a 2cm spacing between the heels with extended legs. Two familiarization trials were performed for each balance task before PRE session data acquisition in order to avoid any learning effect (Cug and Wikstrom, 2014). Displacements of the centre of foot pressure (COP) were recorded at 40Hz. COP surface area (COPS: 90% confidence ellipse), mean COP velocity along the medio-lateral (COPX) and anterior-posterior (COPY) axes were calculated. These COP parameters provide a reliable and valid outcome measure to characterise postural sway and quantify MF-related balance control impairments (Hachard et al., 2020; Ma et al., 2009, Morris and Christie, 2020). The relative difference of each COP parameter (RD =100*[POST-PRE]/PRE) was also calculated.

## 2-5- Statistical analysis

A hierarchical cluster analysis (HCA) was performed to classify participants' behaviour into subgroups with similar characteristics (Watelain et al., 2000; White and McNair, 2002). This analysis is a multivariate statistical method which places individuals into a cluster that contains other individuals with similar characteristics thus enabling the identification of natural groupings that may exist in a population (White and McNair, 2002). Data from the NASA-TLX test, AX-CPT and balance control assessments were independently analysed in order to produce three separated cluster classifications: NASA-TLX clusters, AX-CPT clusters and balance clusters. A normalised principal component analysis (PCA) was processed as a preliminary step to reduce the dimensionality of the data prior to performing the HCA based on the 5 first principal component (PC) scores. Each dimension score and the global score from the NASA-TLX test were used to produce a 19 × 7 input matrix (19 [number of participants] × 7 [input variables]) for the PCA employed to produce the NASA-TLX clusters. Concerning the AX-CPT, PRE and RD values of RT and CR (RT_PRE, RT_RD, CR_PRE and CR_RD) were used to produce a 19 × 4 input matrix for the PCA employed to produce the AX-CPT clusters while taking into account the initial levels and evolutions of participants. In the same manner, each balance task (EO and EC) was independently analysed in a specific PCA while using PRE and RD values of each COP parameter (COPS_PRE, COPX_PRE, COPY_PRE, COPS_RD, COPX_RD and COPY_RD) to produce a 19 × 6 input matrix for the PCA employed to produce the balance clusters.

In order to examine the relationships that could exist between behaviour typologies from subjective workload, cognitive performance and balance control data, the cluster number of each participant was assigned as a categorical variable in each type of cluster (*i.e.* NASA-TLX clusters, AX-CPT clusters and balance clusters). Then, contingency tables between cognitive performance / subjective workload and balance control categorical variables were constructed. The link between these variables were then tested with the Fisher's exact test. All analyses were performed with the R statistical software R, especially with the FactomineR package (Lê et al., 2008). The level of significance was set at $p < 0.05$.

**3- Results**

3-1- Hierarchical cluster analysis

Due to a technical problem, only 19 of the 20 subjects were considered for the analysis. PCA performed on NASA-TLX data resulted in two components that explained 59.0% of the total variance of the original dataset. PC1 accounted for 40.1% of the total variance and was loaded with three variables, EFFORT, FL and GS, which were positively correlated. PC2 explained 18.9% of the total variance and was mainly loaded with three variables, PERF and TD, which were positively correlated, and PD which was acting in the opposite direction (Figure 2a). There was a high variability between individuals' subjective responses to the prolonged cognitive task and the clustering analysis identified three clusters (Figure 2b), the characteristics of which are detailed in Table 1. About 60% of participants showed signs of moderate subjective mental workload (individuals from C2), whereas about 20% of participants showed high subjective mental workload (individuals from C3) and about 20% of participants showed low subjective mental workload (individuals from C1).

PCA performed on the AX-CPT resulted in two components that explained 65.3% of the total variance of the original dataset. PC1 accounted for 35.3% of the total variance and was loaded with RT_PRE and RT_RD, which were negatively correlated. PC2 explained 30.0% of the total variance and was loaded with CR_PRE and CR_RD, which were negatively correlated (Figure 3a). As illustrated on the individuals' factor map (Figure 3b), the HCA revealed two subgroups of individuals, who either presented signs of cognitive performance impairment (individuals from C1, n = 11) or not (individuals from C2; n = 8). The characteristics of these clusters are detailed in Table 2.

PCA performed on COP parameters in the EO balance task resulted in two components that explained 58.1% of the total variance of the original dataset. PC1 accounted for 37.1% of the total variance and was loaded with three variables, COPX_PRE and COPY_PRE, which were positively correlated and COPY_RD, which was acting in the opposite direction. PC2 explained 21.0% of the total variance and was mainly loaded with two variables, COPX_RD and COPY_PRE, which were positively correlated (Figure 4a). As illustrated on figure 4b, there was a high inter-subject variability and three clusters was identified by the HCA, the characteristics of which are detailed in Table 3. Balance control was impaired by MF in the two individuals from cluster 1 (higher COPX_RD and COPY_RD values than the whole group) and in the seven individuals from cluster 2 (higher COPS_RD), who swayed less than other participants at baseline (lower values of COPS_PRE). Individuals from cluster 3 (n = 10) swayed more than other participants at baseline (higher values of COPS_PRE), but they displayed low COPS_RD (unchanged sway due to MF).

PCA performed on COP parameters in the EC balance task resulted in two components that explained 67.9% of the total variance of the original dataset. PC1 accounted for 47.7% of the total variance and was loaded with four variables, COPX_RD and COPY_RD which were positively correlated, and COPX_PRE and COPY_PRE, which were acting in the opposite direction. PC2 explained 20.2% of the total variance and was mainly loaded with COPS_PRE only (Figure 5a). As illustrated on the individuals' factor map (Figure 5b), three clusters were identified by the HCA, the characteristics of which are detailed in Table 3. Balance control of the two individuals in cluster 1, who had high values of COPX_PRE, was not affected by MF (negative COPX_RD and COPY_RD values). Balance control of individuals from cluster 2 (n = 11), who had high values of COPS_PRE and low values of COPS_RD, was not also strongly affected by MF. On the contrary, in individuals from cluster 3 ( n = 6) who had had low values of COPY_PRE and COPX_PRE, there was an increased sway due to MF characterised by high COPS_RD, COPX_RD and COPY_RD.

3-2- Association between clusters

Relying on clusters from the HCA, contingency tables between NASA-TLX clusters, AX-CPT clusters and balance clusters were constructed by assigning for each subject the cluster number as a categorical variable. The Fisher's exact test showed that there was a significant association between balance clusters in the EO task and NASA-TLX clusters ($p$ = 0.02794) and between balance clusters in the EO task and AX-CPT clusters ($p$ = 0.02671). There was no significant association between NASA-TLX clusters and AX-CPT clusters. There was also no significant association between balance clusters and NASA-TLX / AX-CPT clusters in the EC balance task. Figure 6 presents the percentage of association between balance clusters in the EO task, NASA-TLX clusters and AX-CPT clusters. The two subjects from C1 balance (individuals with higher COPX_RD and COPY_RD values), belong to C3 NASA-TLX (individuals with higher values of TD, FL and GS dimensions) and are also included in C1 AX-CPT (which encompasses participants with negative CR_RD, *i.e.* decrease in CR, and positive RT_RD, *i.e.* increase in RT). Most of individuals from C2 balance (with lower values of COPS_PRE and higher COPS_RD) are included in C2 NASA-TLX (individuals showing moderate signs of subjective mental workload) and in C1 AX-CPT. Finally, subjects from C3 balance (individuals with higher values of COPS_PRE and lower COPS_RD) are mainly divided between C1 NASA-TLX (individuals with lower values of GS, MD and EFFORT) and C2 NASA-TLX, and they are also mainly included in C2 AX-CPT (which encompasses participants with almost null CR_RD and negative RT_RD - *i.e.* decrease in RT).

## 4- Discussion

Our previous study performed on aggregate data showed that the fulfilment of the AX-CPT generated a subjective and objective state of MF with a deterioration in balance control (Hachard et al., 2020). Further studies recently confirmed these findings (Morris and Christie, 2020; Tassignon et al., 2020; Varas-Diaz et al., 2020). In the current paper, we reanalysed our previously published data (Hachard et al., 2020) to investigate differences in individual responses to the 90-minute AX-CPT in order to provide new insights about the influence of mental fatigue on balance control. The cluster analysis revealed that the 90-minute AX-CPT induced various levels of MF and balance impairments within a group of healthy young subjects. There was a relationship between the level of MF (characterized by NASA-TLX and AX-CPT responses) and the degree of balance control disturbance only when subjects stood quietly with the eyes open.

Results from the PCA showed that there was a high variability between individuals' responses following the completion of the AX-CPT. The HCA also provided evidence that this variability did not present a random structure but was structured in clusters of participants with similar characteristics. These results show that the AX-CPT produced different levels of MF and different MF-related balance control impairments within a homogeneous population of healthy young subjects. These findings indicate the existence of marked individual differences in MF vulnerability similar to those reported in studies about the effects of fatigue induced by sleep deprivation, with vulnerable and resistant individuals to MF (Tkachenko and Dinges, 2018; Van Dogen et al., 2004; Van Dogen, 2005). Explanations about the sources of this inter-individual variability can be proposed by examining the association between NASA-TLX, AX-CPT and balance clusters.

The absence of significant association between NASA-TLX clusters and AX-CPT clusters is consistent with other previous studies which described a dissociation between subjective ratings of MF and objective assessments of cognitive performance outcomes (Dennis et al., 2017; Rupp et al., 2010; Verschuren et al., 2020). MF is a complex and multidimensional phenomenon (Verschuren et al., 2020) and participants' ranking in terms of vulnerability or resistance to MF varies depending on the task or measure (Dennis et al., 2017). In the present study, this absence of association between NASA-TLX clusters and AX-CPT clusters can be related to the limited effectiveness of the AX-CPT for inducing a homogeneous level of MF among the investigated sample. Individuals from C2 AX-CPT did not present any objective sign of mental fatigue. The decrease in RT in the absence of decrease in CR between the beginning and the end of the AX-CPT observed in these individuals rather characterise a habituation/practice effect than a cognitive impairment. Delliaux et al. (2019) emphasised that RT could decrease while performing a long duration cognitive task due to a habituation phenomenon. O'Keeffe et al. (2020) suggested that, despite its widespread use in MF protocols, the 90-minute AX-CPT did not always produce high level of mental fatigue per se, but could also induce a state of under-arousal post-test (*i.e.* sleepiness, boredom). In some subjects, the 90-minute AX-CPT might have induced a relatively high subjective mental workload due to struggle against boredom and sleepiness without impairing cognitive processes (as observed in individuals from C2 AX-CPT) thus explaining the absence of significant association between NASA-TLX clusters and AX-CPT clusters.

The Fisher's exact test showed that there was a significant association between vulnerability in subjective reports (*i.e.* NASA-TLX clusters) and balance control impairments (*i.e.* balance clusters) and a significant association between cognitive impairments (AX-CPT clusters) and balance control impairments when the visual cues were available. In this task, balance control was impaired only in participants who presented the highest scores at the NASA-TLX test and in participants who experienced the most degraded performance at the AX-CPT. Conversely, balance control was not impaired in participants with low NASA-TLX scores and those with non-degraded AX-CPT performance. When focusing on the features of the most vulnerable participants, they were those who swayed the least at baseline in the EO balance task. These participants were able to face the steadiness requirements at baseline in a very efficient way, likely due to a high engagement of visual attention associated to a high neural activity in the brain areas that regulate attentional processes (Bonnet, 2016; Mihara et al., 2008; Varghese et al., 2015). Even though CR and RT at the beginning of the MF intervention were not variables that significantly discriminate against cluster membership, it could be noted that vulnerable subjects had lower RT values at baseline than resistant subjects. This feature could also reflect a greater engagement of visual attention during the AX-CPT, which may also have led to visual fatigue (Dillon and Emurian, 1996). Hence, one can hypothesize that inter-individual variability could be related to a difference in the level of engagement of visual attention in both the 90-minute AX-CPT and balance tasks.

Resistant participants might have engaged less attentional resources in the AX-CPT while adopting a more automatic detection of target letters, which enabled them to optimize task effectuation after a certain amount of time on task and to implement a habituation process characterized by a decrease in RT without any concomitant decrease in CR (Csathó et al., 2012). O'Keeffe et al. (2020) recently showed that the 90-minute AX-CPT did not always allow participants to be at an optimum state of arousal during the test, which could induce variability in attention engagement, cognitive effort and MF. Our results seem to confirm these findings by indicating that MF would not be achieved by the 90-minute AX-CPT in all subjects. As suggested by Van Dogen (2005) to explain differential vulnerability to sleep deprivation, it can also be hypothesized that resistant individuals to MF have higher cognitive reserve (Stern, 2002, 2009) than vulnerable individuals. Cognitive reserve refers to individual differences in the ability to cope with behavioral and cognitive deterioration due to task demand, neurodegenerative disorders or brain injury, which may allow some individuals to be more resistant than others (Stern, 2002, 2009). Neural implementation of cognitive reserve can stem from a more efficient activation of given brain networks (*i.e.* inter-individual differences in cognitive processing) or an enhanced ability to implement compensatory neural activity by recruiting alternate brain networks (Stern, 2002, 2009).

Given that the association between the level of MF and the degree of balance disturbance was only significant in the EO balance task, one can suggest that inter-individual differences in vulnerability to MF could stem from differences in preferred modes of spatial referencing, *i.e.* dependence-independence on visual field information for spatial orientation. Visual field dependence-independence strongly interacts with the visual contribution to postural control: field-dependent individuals use predominantly visual cues, while field-independent individuals rely rather on non-visual cues, such as vestibular, tactile and proprioceptive cues to perceive spatial orientation and control their balance (Isableu et al., 1997; Streepey et al., 2007). Our results showed that participants who swayed the least at baseline when visual cues were available were also the most vulnerable individuals to MF. One could assume that these participants were initially the most field-dependent, which enabled them to exploit visual cues more efficiently to control their balance with open eyes. A visual reweighting with a reduced contribution of visual input to balance control is classically observed in field-dependent subjects when they are exposed to strong optokinetic visual stimulations (Mahboobin et al., 2005; Pavlou et al., 2011; Scotto Di Cesare, 2015). Hence, the sustained visual stimulation imposed by the completion of the 90-minute AX-CPT, with a potential task-related visual fatigue phenomenon, might have forced field-dependent participants to engage a sensory reweighting process and to prioritize vestibular and somatosensory inputs when performing the EO balance task, thus leading to higher postural disturbances than in field-independent participants who are accustomed to exploit non-visual inputs to control their balance. Future studies should be conducted to accurately identify the factors that underlie individual differences in MF vulnerability, in particular by assessing participants' attentional engagement and their dependency on visual information, but also by examining potential manifestations of visual fatigue. Neuroimaging techniques could also be implemented in order to identify potential brain compensatory mechanisms or difference in baseline brain activity associated to cognitive reserve.

Limitations of the current research are acknowledge. The first limitation is the small sample size. Further studies need to be conducted with larger sample size to reach higher statistical power and confirm our findings. Secondly, we did not assess the participants' sitting posture when they performed the 90-minute AX-CPT. In our previous study (Hachard et al., 2020), we showed that balance control was impaired in a control condition which consisted of watching a 90-minute documentary. This impairment of balance control was attributed to a deleterious effect of prolonged sitting on balance control since the control condition did not engender any subjective and objective state of MF. Prolonged sitting can indeed induce a creep of passive structures, which adversely affects sensorimotor trunk control mechanisms by reducing back muscles spindle sensitivity (Howarth et al. 2013, Kastelic et al. 2018, Sanchez-Zuriaga et al. 2010), with potential negative effects in activities performed following extended bouts of sitting such as standing balance. These alterations in neuromuscular control due to prolonged sitting are more pronounced with slumped sitting postures (Howarth et al., 2013). Participants in the present study adopted an upright sitting posture in order to stay focused on the screen while performing the AX-CPT with their hands on the keyboard, which would have limited deleterious effects of prolonged sitting on balance control during subsequent standing. Nevertheless, we cannot exclude that some subjects had adopted a slumped sitting posture with a potentially greater adverse effect on balance control, which may have influenced the distribution of participants between balance clusters. Thirdly, although participants were instructed to maintain regular sleep habits before assessment sessions, we did not monitor sleep duration and quality prior to the tests. Quality and quantity of sleep can affect cognitive performance and motor control performance such as balance control (Ma et al., 2019; Rupp et al., 2010, Van Dongen, 2010, Van Dongen et al., 2004). Hence, individual differences in sleep duration and quality could also influence the distribution of participants between balance clusters.

Future studies about the influence of MF induced by a prolonged cognitive task should control participants' sitting posture during the prolonged cognitive task and sleep quantity and quality before assessment sessions to prevent any confounding bias. Sitting posture can be assessed with kinematic and/or electromyographic analyses and sleep quality/quantity can be easily monitored by self-rated questionnaires (e.g. Pittsburgh Sleep Quality Index) and/or objective sleep parameters (e.g. sleep actigraphy).

## 5- Conclusion

The present study provides new insights about the influence of MF on balance control by analysing inter-individual differences in participants' responses to MF with a hierarchical cluster analysis. It shows that, among a sample of healthy young subjects, the completion of the same 90-minute AX-CPT cognitive task induced different levels of MF that affect balance control differently according to the sensory context. The relationship between the level of MF and the degree of balance control disturbance was significant only when participants stood with their eyes open. This result suggests that that inter-individual differences in MF-related balance control disturbances could stem from differences between participants in the level of engagement of visual attention and/or from differences in visual dependency for balance control. This study emphasizes the necessity to address the issue of inter-individual variability when studying the effects of MF. Although taking into account inter-individual variability may complicate the analysis of data from MF studies because of the need to implement adequate methods (e.g., clustering analyses), investigations based solely on group ensemble average methods, as usually employed in the field of fatigue research, do not provide an accurate representation of individual reactions and may produce misleading conclusions.

**Acknowledgements** We would like to thank all the participants involved in this study.

**Author contributions** Conceptualization: FN, BH, HC, NB, TP; methodology: FN, BH, HC, NB; data curation: FN; formal analysis: FN, BH; software: FN, NB; writing-original draft preparation: FN; writing-review and editing: BH, HC, NB, TP; supervision: TP.

**Funding** This research did not receive any specific grant from funding agencies in the public, commercial, or not-for-profit sectors.

**Compliance with ethical standards**

**Conflict of interest** The authors declare that they have no conflict of interest.

**Informed consent** Informed consent was obtained from all individual participants included in the study.

**Data availability** The datasets generated during and/or analysed during the current study are available from the corresponding author on reasonable request.

**Caption of the figures**

**Figure 1.** Illustration of the experimental protocol.

MF intervention consisted of performing a 90-min continuous demanding cognitive task (AX-CPT). Balance control and performance during the AX-CPT (CR: percentage of correct responses; RT: reaction time) were assessed at baseline (PRE) and at the end of the MF intervention (POST). The relative differences (RD =100*[POST-PRE]/PRE) was calculated in order to characterizes the evolution of the variables between PRE and POST measures. Subjective mental workload was also assessed with the NASA-TLX test at the end of the MF intervention.

**Figure 2**. Variables' factor map (a) of the PCA applied on NASA-TLX data and hierarchical clustering on the factor map (b)

MD: mental demand; PD: physical demand; TD: temporal demand; FL: frustration level; EFFORT: effort; PERF: performance; GS: global score.

**Figure 3**. Variables' factor map (a) of the PCA applied on the AX-CPT and hierarchical clustering on the factor map (b)

CR_PRE and RT_PRE: percentage of correct responses (CR) and reaction time of correct responses (RT) obtained during the first 10-min period of the AX-CPT; CR_RD and RT_RD: relative difference of CR and RT between the first and last 10-min periods of the AX-CPT (RD = [POST-PRE])/PRE*100).

**Figure 4**. Variables' factor map (a) of the PCA applied on the COP parameters and hierarchical clustering on the factor map (b) in the EO balance task

COPS_PRE: COP surface area at baseline; COPX_PRE: mean COP velocity along the medio-lateral axis at baseline; COPY_PRE: mean COP velocity along the antero-posterior axis at baseline; COPS_RD, COPX_RD and COPY_RD: relative difference of COPS, COPX and COPY variables (RD = [POST-PRE])/PRE*100)

**Figure 5**. Variables' factor map (a) of the PCA applied on the COP parameters and hierarchical clustering on the factor map (b) in the EC balance task

COPS_PRE: COP surface area at baseline; COPX_PRE: mean COP velocity along the medio-lateral axis at baseline; COPY_PRE: mean COP velocity along the antero-posterior axis at baseline; COPS_RD, COPX_RD and COPY_RD: relative difference of COPS, COPX and COPY variables (RD = [POST-PRE])/PRE*100)

**Figure 6**. Percentage of association between balance clusters in the EO task, NASA-TLX clusters and AX-CPT clusters

Table 1. Results of NASA-TLX in the whole sample and in each cluster

|  | Whole sample (n=19) | C1 (n=4) | C2 (n=11) | C3 (n=4) |
|---|---|---|---|---|
| MD | 64.4 (26.9) | 22.1 (10.2)* | 74.2 (17.0) | 79.8 (13.0) |
| PD | 37.1 (33.1) | 29.9 (32.6) | 31.7 (30.9) | 59.2 (33.4) |
| TD | 45.2 (29.4) | 36.9 (27.1) | 35.9 (26.3) | 78.9 (13.0)* |
| PERF | 62.6 (19.0) | 76.3 (14.9) | 52.3 (16.6)* | 77.3 (9.0) |
| EFFORT | 67.6 (22.5) | 43.8 (19.7)* | 71.3 (21.5) | 81.4 (5.7) |
| FL | 38.0 (26.6) | 17.3 (6.9) | 35.7 (26.2) | 64.8 (16.3)* |
| GS | 59.7 (14.3) | 44.5 (13.4)* | 57.8 (6.3) | 80.0 (5.9)* |

Data are expressed as mean (SD). MD: mental demand; PD: physical demand; TD: temporal demand; PERF: performance; EFFORT: effort; FL: frustration level; GS: global score; C1: cluster 1; C2: cluster 2; C3: cluster 3. The number in the brackets indicates the number of the subjects that are included in a specific cluster. * indicates variables with values that differ significantly from those of the whole sample ($p < 0.05$) and that best discriminate the corresponding cluster

Table 2. Results of AX-CPT in the whole sample and in each cluster

|  | Whole sample (n=19) | C1 (n=11) | C2 (n=8) |
|---|---|---|---|
| CR_PRE (%) | 96.8 (1.8) | 97.4 (1.9) | 96.0 (1.3) |
| CR_RD (%) | -3.1 (4.9) | -5.2 (5.4)* | -0.3 (2.1)* |
| RT_PRE (ms) | 394.8 (61.1) | 371.8 (42.9) | 426.4 (70.5) |
| RT_RD (%) | -4.4 (13.8) | 2.4 (13.5)* | -13.7 (7.9)* |

Data are expressed as mean (SD). CR_PRE and RT_PRE: percentage of correct responses and reaction time obtained during the first 10-min period of the AX-CPT; CR_RD and RT_RD: relative difference of CR and RT between the first and last (POST) 10-min periods of the AX-CPT [RD =100*(POST-PRE)/PRE]; C1: cluster 1; C2: cluster 2. * indicates variables with values that differ significantly from those of the whole sample ($p < 0.05$) and that best discriminate the corresponding cluster

Table 3. COP parameters from balance assessments in the whole sample and in each cluster

| | | Whole sample | C1 (n=2) | C2 (n=7) | C3 (n=10) |
|---|---|---|---|---|---|
| EO | COPS_PRE | 238 (113.6) | 171.4 (8.3) | 164.3 (72.6)* | 302.8 (111.4)* |
| | COPS_RD | 27.0 (56.5) | 54.7 (69.1) | 65.6 (56.6)* | -5.5 (34.9)* |
| | COPX_PRE | 6.6 (1.2) | 5.2 (0.2) | 5.9 (0.7) | 7.3 (1.0)* |
| | COPX_RD | -1.2 (26.2) | 54.7 (40.2)* | -5.1 (17.1) | -9.8 (14.6) |
| | COPY_PRE | 5.0 (1.2) | 4.7 (2.6) | 4.4 (0.6) | 5.5 (1.1) |
| | COPY_RD | 3.3 (27.2) | 54.1 (54.5)* | -5.4 (15.3) | -0.7 (18.6) |
| | | Whole sample | C1 (n=2) | C2 (n=11) | C3 (n=6) |
| EC | COPS_PRE | 367.1 (174.3) | 298.3 (34.4) | 441.5 (191.8)* | 253.8 (77.9) |
| | COPS_RD | 32.4 (51.9) | 39.6 (54.8) | 6.0 (40.7)* | 78.3 (41.2)* |
| | COPX_PRE | 9.4 (2.4) | 14.7 (0.3)* | 9.4 (1.7) | 7.7 (0.7)* |
| | COPX_RD | 18.9 (25.8) | -20.6 (14.1)* | 15.1 (18.3) | 39.1 (23.2)* |
| | COPY_PRE | 8.0 (2.3) | 11.4 (2.4)* | 8.6 (1.8) | 5.9 (1.1)* |
| | COPY_RD | 27.4 (30.3) | -20.3 (14.1)* | 20.3 (16.5) | 56.4 (26.3)* |

Data are expressed as mean (SD). EO: eyes open; EC eyes closed; COPS_PRE, CPOX_PRE and COPY_PRE: COP surface area, mean COP velocity along the medio-lateral and anterior-posterior axes before performing the AX-CPT; COPS_RD, CPOX_RD and COPY_RD: relative difference of COPS, COPX and COPY PRE and POST AX-CPT measurements [RD =100*(POST-PRE)/PRE]; C1: cluster 1; C2: cluster 2; C3: cluster 3. * indicates variables with values that differ significantly from those of the whole sample ($p < 0.05$) and that best discriminate the corresponding cluster